\documentclass[prd,twocolumn,tightenlines,eqsecnum,floats,aps,amsmath,amssymb,nofootinbib,shownopacs,floatfix]{revtex4}

\usepackage{graphicx, wrapfig}
\usepackage{amssymb}
\usepackage{dcolumn}
\usepackage{bm}
\usepackage{physics}
\usepackage[toc,page]{appendix}
\usepackage{color}
\usepackage{bbm}
\usepackage{mathptmx}
\usepackage{amsfonts}
\usepackage{graphicx}
\usepackage{bbm}
\usepackage{color}
\usepackage{pifont}
\usepackage{enumerate}
\usepackage{subfigure}
\usepackage{color}
\newcommand*{\rom}[1]{\expandafter\@slowromancap\romannumeral #1@}
\makeatother
\usepackage{mathptmx}
\usepackage{mathrsfs}
\usepackage{mathtools}
\usepackage{amsmath}
\usepackage{amsfonts}
\usepackage{amssymb}
\usepackage{mathdots}
\usepackage{bbold}
\usepackage{isomath}
\usepackage{mathrsfs}
\usepackage{amsthm} 
\usepackage{graphicx}
\usepackage{bbm}
\usepackage{color}
\usepackage{pifont}
\usepackage{enumerate}
\usepackage{amsthm}
\usepackage{braket}
\theoremstyle{plain}
\theoremstyle{definition}
\usepackage{mathtools}
\usepackage{color}
\usepackage{mathrsfs}
\usepackage{subfigure}
\makeatletter
\newsavebox{\@brx}
\newcommand{\llangle}[1][]{\savebox{\@brx}{\(\m@th{#1\langle}\)}%
  \mathopen{\copy\@brx\kern-0.5\wd\@brx\usebox{\@brx}}}
\newcommand{\rrangle}[1][]{\savebox{\@brx}{\(\m@th{#1\rangle}\)}%
  \mathclose{\copy\@brx\kern-0.5\wd\@brx\usebox{\@brx}}}
\makeatother

\newcommand{\be}{\begin{equation}}
\newcommand{\ee}{\end{equation}}
\newcommand{\ba}{\begin{eqnarray}}
\newcommand{\ea}{\end{eqnarray}}

\newcommand{\upspin}{\ket{\uparrow}}
\newcommand{\downspin}{\ket{\downarrow}}
\newcommand{\nbf}{{\hat{\bf n}}}
\newcommand{\nbfup}{{\nbf}_\uparrow}
\newcommand{\nbfdn}{{\nbf}_\downarrow}
\newcommand{\aup}{{{\hat{a}_\uparrow}}}
\newcommand{\adn}{{{\hat{a}_\downarrow}}}
\newcommand{\hata}{\hat{a}}
\newcommand{\hatS}{\hat{S}}
\newcommand{\hats}{\hat{s}}
\newcommand{\hatq}{\hat{q}}
\newcommand{\hatX}{\hat{X}}
\newcommand{\hatP}{\hat{P}}
\newcommand{\hatH}{\hat{H}}

\newcommand{\eref}[1]{{Eq.~(\ref{#1})}}

\begin{document}

\preprint{APS/123-QED}

\title{Towards quantum simulation of spin systems using continuous variable quantum devices}

\author{Razieh Annabestani}
\email{rannabes@uwaterloo.ca}
\affiliation{Transformative Quantum Technologies, University of Waterloo, Canada}

\author{Brajesh Gupt}
 \email{bgupt@tacc.utexas.edu}
 \affiliation{Texas Advanced Computing Center, The University of Texas at Austin, TX USA}

\author{Bhaskar Roy Bardhan}
 \email{bhasiitg@gmail.com}


\begin{abstract}
We study Bosonic representation of spin Ising model with the application of
simulating two level systems using continuous variable quantum processors. We
decompose the time evolution of spin systems into a sequence of continuous
variable logical gates and analyze their structure. We provide an estimation of
quantum circuit scaling with the size of the spin lattice system. Furthermore,
we discuss the possibility of using a Gaussian Boson sampling device to estimate
the ground state energy of Ising Hamiltonian. The result has potential application in
developing hybrid classical-quantum algorithms such as continuous variable
version of variational quantum eigensolver. 
\end{abstract}

\pacs{Valid PACS appear here}

\maketitle

\section{Introduction}
\label{sec_1}
It is known that simulating a quantum system is a computationally hard problem
\cite{Lloyd96,AspuruGuzik2005SimulatedQC,WhitfieldetalQuantumSimulation,Babbush_2017,Georgescu:2013oza}.
The difficulty stems from the exponential explosion in the computational
resources needed for the simulation, leading to strong limitations on the size of quantum systems that can be simulated using modern supercomputers.\footnote{For certain problems the exponential scaling can be
circumvented by resorting to approximate methods such as Monte-Carlo methods.
However, due to unavailability of such approximation in general, quantum
simulation remains a computationally hard problem, even for modern
supercomputers.} As observed by Feynman in 1982, a quantum simulator can resolve
the problem of exponential explosion in the required computational
resources and perform the simulation in polynomial time
\cite{feynman1982simulating}.  Over the past 40 years, there have
been extensive research and tremendous efforts to build quantum devices as
well as quantum algorithms that can realize quantum
simulations
\cite{Lloyd96,PhysRevLett.79.2586,PhysRevLett.83.5162,Zalka_1998,Kassal_2008,AspuruGuzik2005SimulatedQC,Brown_2010,WhitfieldetalQuantumSimulation,Babbush_2017,Georgescu:2013oza}. 

In a typical classical simulation, one begins with the Hamiltonian that
describes the energy of the physical system and solves the dynamical equations
to obtain the time evolution of the system of interest. That information is used
to compute the ground state energy or other relevant properties of the system.
In a quantum simulation, however, the Hamiltonian of the physical system is
first mapped into the degrees of freedom of another quantum system that we can
fully control and manipulate to extract information
\cite{Lloyd96,AspuruGuzik2005SimulatedQC,Georgescu:2013oza}. If the quantum
information is transformed and stored faithfully, the quantum simulator emulates the dynamic of the physical system of interest. By applying specific quantum operations and measurements on the quantum processor, the time evolution can be studied and
the expectation value of the Hamiltonian for a given state can be computed.
\cite{Lloyd96,PhysRevLett.79.2586,PhysRevLett.83.5162,Zalka_1998,Kassal_2008,AspuruGuzik2005SimulatedQC,Brown_2010,WhitfieldetalQuantumSimulation,Babbush_2017,Georgescu:2013oza}.

One of the most important problems in quantum simulation is that of simulating
spin systems. In particular, Ising spin systems have been widely used to
understand phase transition in magnetic systems and to study lattice gas and spin glass systems
\cite{Brush_67_IsingRMP,stanley1987introduction,binney1992theory}. It has been
shown that the classical simulation of an Ising model is an NP-complete problem
\cite{Barahona_1982} and many NP-complete problems can be mapped to Ising model
\cite{Lucas_2014}. In particular, the Hamiltonian of a molecular system can be
mapped to an Ising model \cite{Xia2017ElectronicSC}. Hence, solving the Ising
model on a quantum simulator paves the way to solving many complex problems of
scientific and industrial interests.

Continuous variable (CV) quantum computing is a branch of quantum information processing where 
information is stored in quantum observables with continuous spectrum--such as positions and 
momenta operators. In the conventional qubit model, information is stored in 
discrete levels of a quantum system such as spins or two levels system of an atom or a 
superconducting system \cite{RevModPhys.84.621,RevModPhys.77.513}. In analogy to qubit, the 
building block of a CV quantum processor are photonic modes whose states are described in an 
infinite dimensional Hilbert space. Quantum information is encoded in the amplitude and
phase of electromagnetic waves or light.  The associated CV quantum gates are implemented by 
passive optical elements such as phase shifters, beam splitters,
inteferometers or active interaction with non-linear crystals. 
\cite{PhysRevLett.82.1784,PhysRevA.93.052304}. CV quantum computing has been shown 
to be equivalent to qubit based quantum computing \cite{nielsen_chuang_2010}.

In this work, we discuss the simulation of Ising spin model on a continuous
variable (CV) quantum processor (for simulations using qubit based quantum
computers and quantum annealers see
\cite{lidar_ising,Santoro_2002,Boixo_2013,Boixo_2014,Johnson2011Quantum,Whitfield_2012,Biamonte2008RealizableHF,Cervera_Lierta_2018}
and references therein.). Through Jordon-Schwinger transform, we first map
Edward-Anderson and Heisenberg Hamiltonian for spin systems to Bosonic degrees
of freedom used in CV quantum computing. This results in quadratic and quartic
interaction terms in the CV quantum modes. We analyze these terms and study CV
gate decomposition of the resulting Hamiltonian in terms of the universal CV gate
set. This will enable us to provide a resource estimate needed to simulate
Ising model on a full-fledged CV quantum computer. While a full fledged quantum 
processor is years away, in recent years, near term noisy quantum devices (NISQ) 
have been used to estimate the expectation value of molecular Hamiltonians. The output 
of the NISQ devices are then fed to classical computers to variationally estimate the ground 
state energy
\cite{Peruzzo_2014,McClean_2016,PhysRevX.6.031007,Linke_2017,Kandala_2017,PhysRevLett.120.210501,McCaskey2019QuantumCA}. We propose to
estimate the ground state energy of Ising model using Gaussian Boson sampling
(GBS) device, a near term continuous variable quantum device.  

The paper is organized as follows. Section \ref{sec_2} gives a brief
introduction to the Ising model under consideration and presents mapping of the
Ising Hamiltonian to CV degrees of freedom. In section \ref{sec_3}, we discuss
various terms in the CV Hamiltonian, their decomposition in to universe CV
quantum gates and a resource count to simulate the Ising model. In section
\ref{sec_4}, we discuss estimation of quartic terms in Hamiltonian involving
four mode optical correlation using a Gaussian Boson sampling device. We
summarize our result in section \ref{sec_5}.

\section{Bosonic Hamiltonian for spin system}
\label{sec_2}
\subsection{Spin Hamiltonian}
Spin systems are usually defined on a $d-$dimensional lattice with $N$ number of
sites each being occupied by spin $S_i$. There are two main components to the
energy of the spin system: (i) due to the interaction between spins with their
nearest neighbors and (ii) due to interaction of the individual spins with an
external magnetic field. Thus, In the presence of an external magnetic field, $\vec{B} =
B_{0}\ \hat{z}$, the energy of such a system is given by the following
Hamiltonian:
\begin{equation}
\hat{H}_{\text{Ising}} = - \sum\limits_{k,l =1, k<l}^{N}\  J_{kl}^{z}~\hat{s}_k~\hat{s}_l
 - B_{0}\ \sum \limits_{k=1}^{N} \ \hat{s}_k.
\label{EdwardAndersonH}
\end{equation}
where $J^{z}_{kl}$ are the coupling constants between spins at the
$k^{\text{th}}$ and the $l^{\text{th}}$ sites and $\hat{s}_i$ denotes the spin operator 
(with eigenvalues $s_i=\{+1, -1\}$) at $i^{\rm th}$ site. The first term of the Hamiltonian
represent the internal spin-spin interaction and the second term represents the
energy due interaction of an individual spin with the external magnetic field,
taken to be homogeneous accros the lattice. Ising model is one of the most widely studied spin system \cite{Lenz_1920,Ising_1925,Brush_67_IsingRMP}, in which each spins
is given by a two state system, given by $\{+1,-1\}$. In a magnetic system, the
binary state represents the two configurations, namely aligned
$\upspin$ and anti-aligned $\downspin$ states with an external
magnetic field.

The Ising model was formulated by Wilhelm Lenz and Ernst Ising to
describe the behavior of anti-ferromagnetism in a spin lattice
\cite{Lenz_1920,Ising_1925,Brush_67_IsingRMP}. This simple and
yet rich model has become a very popular model for representing distinct
physical systems due to various interpretation of ``spin''. For instance, one
interpretation of ``spin'' can be the presence $(s= + 1)$ or the absence $(s=
-1)$ of a molecule in a certain cell of a ``lattice gas'', and so, the model can
be used for studying the critical behavior of a fluid system
\cite{Brush_67_IsingRMP,stanley1987introduction,binney1992theory}. Since the
first application of Ising model to magnetic and molecular systems, 
it has been used to describe different physical systems in various fields
ranging from physics, chemistry, biology to artificial intelligence.
Finding the ground state of Ising model is an NP hard
problem \cite{Barahona_1982} that leads to mapping of many NP hard problems to
the Ising model \cite{Lucas_2014}. This puts Ising model as one of the most
important cornerstones of quantum simulation. A successful quantum simulation
of the Ising model would lead the way to solving many interesting problems, hence
potentially showcasing quantum advantage over the state-of-the-art classical
simulation methods.

The Ising model (or spin systems in general) can be considered as a special
class of Heisenberg models which, in addition to the spin-spin interaction
in $z$-direction ($J^z$) considered in the first term of
\eref{EdwardAndersonH}, also has transverse interactions in $x$ and $y$
directions. The Hamiltonian for a such a system can be written as:

\begin{equation}
\hat{H} = \hat{H}_{\rm trans} + \hat{H}_{\rm Ising},
\end{equation}
where
\begin{equation}
\hat{H}_{\rm trans} = - \sum\limits_{k,l =1, k<l}^{N}\  \left(
			J_{kl}^{x}~\hat{s}_k^x~\hat{s}_l^x 
            + J_{kl}^{y}~\hat{s}_k^y~\hat{s}_l^y 
		\right).
\end{equation}
Therefore the Heisenberg Hamiltonian takes the following form:
\begin{equation}
\hat{H} = - \sum\limits_{k,l =1, k<l}^{N}\  \left(
			J_{kl}^{x}~\hat{s}_k^x~\hat{s}_l^x 
            + J_{kl}^{y}~\hat{s}_k^y~\hat{s}_l^y 
            + J_{kl}^{z}~\hat{s}_k^z~\hat{s}_l^z 
		\right)
		- B_{0}\ \sum \limits_{k=1}^{N} \hat{s}_k^z.
\label{HeisenH}
\end{equation}
In this model, $J^{\alpha}_{kl}$ with $\alpha=\{x,y,z\}$ respectively
are $x$, $y$ and $z$ components of the coupling constants between spins at the
$k^{\text{th}}$ and $l^{\text{th}}$ site.  The above model assumes that the
external field is homogeneous across the lattice and the spin-spin interaction
is isotropic. In the following we will continue to consider binary states for
the spin with eigenvalues $s^{x/y/z}_k=\{+1, -1\}$ 
(corresponding to states \{$\upspin$, $\downspin$\}) and
analyze the Heisenberg Hamiltonian $\hat H$ in \eref{HeisenH}, of which
Ising model is a special case, owing to the isotropy of the spin-spin
interactions. In the next subsection we will transform the spin degrees of
freedom to Bosonic degrees of freedom using Jordon-Schwinger transformation
\cite{jordan1, schwinger2}.

\subsection{Bosonic representation of spin Hamiltonian}

In this paper we are interested in studying the spin systems using a CV
quantum computing device whose physical realization is based on Bosonic degrees
of freedom. Therefore, in order to implement the spin Hamiltonian of
\eref{HeisenH}, we first need to transform the spin degrees of freedom $s_i$
to Bosonic degrees of freedom suitable for a CV quantum device. 
We will consider the standard Jordon-Schwinger
transformation of the angular momentum operators to carry out the transformation
from spin states $s_i$ to the desired Bosonic operators which in our case will be 
based on quantum photonics. We will first transform the spin degrees of freedom
to photonic creation and annihilation operators which are then written in terms
of the continuous degrees of freedom represented by the quantum optical modes.

Let us consider two uncoupled harmonic oscillators described pairs of creation
($a^\dagger_i$) and annihilation $a_i$ operators where the index $i$ represents
the two types of oscillators with $\upspin$ and $\downspin$. The associated
number operators are then defined as follows \cite{schwinger2}:
\be
\nbf_\uparrow = \hat{a}^\dagger_\uparrow \hat{a}_\uparrow, \qquad \nbf_\downarrow =
\hat{a}^\dagger_\downarrow \hat{a}_\downarrow.
\ee
Here, we have assumed the usual commutation relations for $\hat{a}^\dagger_i$ and
$\hat{a}_i$:
\be
\left[\hat{a}_i, \hat{a}_j^\dagger\right] = \delta_{ij}, \quad
\left[\nbf_i, \hat{a}_j^\dagger\right] = \delta_{ij}~]^\dagger_i, \quad
\left[\nbf_i, a_j\right] = -\delta_{ij}~\hata_i.
\ee

The Jordan-Schwinger transformation is a way to write representations of SU(N)
with Bosons \cite{jordan1, schwinger2}.  For SU(2) it is defined by: 
\be
	\hat{S}_\alpha = \frac12 \sum_{i,j} \hat{a}^\dagger_i (\hat{\sigma}_\alpha)_{ij} \hat{a}_j
	\label{S}
\ee
where $\sigma_\alpha$ for $\alpha=x,y,z$ are the Pauli matrices and
$\hat{a}_{i}^{\dagger}, \hat{a}_{i}$ are creation and annihilation operators for
two coupled harmonic oscillators. With the choice of spin basis $\upspin =
\ket{0}$  and $\downspin=\ket{1}$, we can write mix operators defined as
follows:
\ba
\hatS_+  &\equiv& \hatS_x + i~\hatS_y = \hata^\dagger_\uparrow \hata_\downarrow, \nonumber \\
\hatS_-  &\equiv& \hatS_x - i~\hatS_y = \hata^\dagger_\downarrow \hata_\uparrow, \nonumber \\
\hatS_z  &\equiv& \frac{1}{2} \left(\nbf_\uparrow - \nbf_\downarrow\right).
\label{JSmapping}
\ea
Note that the $\hatS_+$ and $\hatS_-$ operators act like spin flip operators, i.e. $\hatS_+$
annihilates a spin of type $\downspin$ and creates a spin of type $\upspin$.
Similarly, $\hatS_-$ annihilates a spin of type $\upspin$ and creates a spin of
type $\downspin$. The operator $\hatS_z$ however simply counts the difference between the
number of spins of two types. All these operators leave the sum of total spins
i.e.\ $\nbf_\uparrow + \nbf_\downarrow$ unchanged. Furthermore, we can define:
\be
\hatS_x \equiv \frac{1}{2} \left(\hatS_+ + \hatS_-\right), \qquad 
\hatS_y \equiv -\frac{i}{2} \left(\hatS_+ - \hatS_-\right).
\ee
It is straightforward to show that the newly defined operators $S_i$ follow the
following commutation relations:
\be
 \left[\hatS_z, \hatS_\pm\right] = \pm \hatS_\pm, \qquad \left[\hatS_+,\hatS_-\right] = 2 \hatS_z.
\ee
The anti-commutation relation between $\hatS_+$ and $\hatS_-$ becomes:
\be
\left\{\hatS_+,\hatS_-\right\} = \nbf_\uparrow + \nbf_\downarrow + 2 \nbf_\uparrow \nbf_\downarrow
\ee
In the model considered here, each lattice site is occupied by a maximum of one
spin-half particle which can be in one of the two states: 
$\upspin = \ket{n_\uparrow=1,n_\downarrow=0}$ or 
$\downspin = \ket{n_\uparrow=0,n_\downarrow=1}$. Therefore, at a given lattice
site, $\nbf_\uparrow + \nbf_\downarrow = 1$ and $\nbf_\uparrow
\nbf_\downarrow=0$. This is a `holonomic' constraint which yields the following
anti-commutation relation for the spin system with spin-half particles:
\be
 \left\{\hatS_+,\hatS_-\right\} = 1.
\label{anticommute}
\ee
This plays a key role in the algebra of the operator and puts $\hatS_+$ and $\hatS_-$
on the similar footings of the ladder operators associated to a single harmonic
oscillator. Furthermore, The holonomic constraint, $\nbf_\uparrow +
\nbf_\downarrow = 1$, projects the infinite dimensional harmonic oscillator
space to a finite dimensional spin space, while making the Jordan-Schwinger
transformation exact on each site.\footnote{Note that at the operator level, the
`holonomic' constraint may not be satisfied in general, in which case the
anti-commutation relation of \eref{anticommute} can be satisfied only under
an approximation.}

\begin{table}
\centering
\caption{The coefficients $C$ and $F$ of quartic terms in \eref{HeisenH_q}.
We have used the following notation to denote various CV terms: $k_1 :=
k\uparrow$, $k_2 := k\downarrow$, $l_1 := l\uparrow$, $l_2 := l\downarrow$}
\label{Table_terms}
\begin{tabular}{ |c| c|c|}
\hline
$q_{k_1} \ q_{k_2} \ q_{l_1} \ q_{l_2}$ & $C$ & $F$ \\
\hline
\hline
 $X_{k_1}\ X_{k_2}\ X_{l_1}\ X_{l_2}$ & +1 & +1  \\
 \hline
 $P_{k_1}\ P_{k_2}\ X_{l_1}\ X_{l_2}$ & +1 & +1  \\
 \hline
 $X_{k_1}\ X_{k_2}\ P_{l_1}\ P_{l_2}$ & +1 & +1  \\
 \hline
 $P_{k_1}\ X_{k_2}\ P_{l_1}\ X_{l_2}$ & +1 & -1 \\
 \hline
 $X_{k_1}\ P_{k_2}\ X_{l_1}\ P_{l_2}$ & +1  & -1 \\
 \hline
  $X_{k_1}\ P_{k_2}\ P_{l_1}\ X_{l_2}$ & -1 & +1\\
 \hline
 $P_{k_1}\ X_{k_2}\ X_{l_1}\ P_{l_2}$ & -1 & +1 \\
 \hline
 $P_{k_1}\ P_{k_2}\ P_{l_1}\ P_{l_2}$ & +1 & +1 \\
 \hline   
\end{tabular}
\end{table}

Using the Jordon-Schwinger mapping i.e. writing ($\hats^x, \hats^y, \hats^z$) in terms of
($\hatS^x,~\hatS^y,~\hatS^z$), $\hatS_+$ and $\hatS_-$, \eref{HeisenH}
can be written as: 
\ba
\hat{H} =  \sum\limits_{k,l =1, k<l}^{N}
       &-&\frac{J_{kl}^x - J_{kl}^y}{4} \left(\hatS_+^k\hatS_+^l + \hatS_-^k\hatS_-^l\right) \nonumber \\
       &-&\frac{J_{kl}^x + J_{kl}^y}{4} \left(\hatS_+^k\hatS_-^l + \hatS_-^k\hatS_+^l\right) \nonumber \\
       &-& J^y_{kl} \hatS^z_k \hatS^z_l ~ - ~ B_0 \sum\limits_{k=1}^{N} \hatS^z_k.
\ea
Now, using the mapping from $S_\pm$ and $S_z$ to creation and annihilation
operators given in \eref{JSmapping}, the Hamiltonian above can be
written as:
\ba
\hat{H} =  \sum\limits_{k,l =1, k<l}^{N}
       \Bigg(&-&\frac{J_{kl}^x + J_{kl}^y}{4} ({\aup}^{\dagger k} {\adn}^k{\aup}^{\dagger l} {\adn}^l
                                               + {\rm h.c.}) \nonumber \\
             &-&\frac{J_{kl}^x - J_{kl}^y}{4} ({\aup}^{\dagger k} {\adn}^k{\adn}^{\dagger l} {\aup}^l
                                               + {\rm h.c.}) \nonumber \\
       &-& J^y_{kl} (\nbfup^k\nbfup^l  - \nbfup^k\nbfdn^l - 
                     \nbfdn^k\nbfup^l  + \nbfdn^k\nbfdn^l) \Bigg) \nonumber \\
       &-& \frac{B_0}{2} \sum\limits_{k=1}^{N} (\nbfup^k - \nbfdn^k),
\label{HeisenH_ada}
\ea
where `h.c.' refers to Hermitian conjugate.

\eref{HeisenH_ada} is the desired Hamiltonian transformed from spin
representation to the Bosonic creation and annihilation operators, which are
discrete Bosonic degrees of freedom. Recall that, in this paper, we are
interested in implementing the Ising model using CV quantum computer. This
requires us to write creation and annihilation operators in term of continuous
degrees of freedom which in this case are the quadrature operators, $q \in
\{X,P\}$:
\ba
 {\hata}&=& {\hatX} + i\ {\hatP} \nonumber \\
 {\hata}^{\dagger}&=& {\hatX} - i\ {\hatP} \nonumber \\
  \nbf &=& {\hatX}^{2} + {\hatP}^2 - \frac{\mathbb{1}}{2},
 \label{Eq_X_P}
\ea
where the standard commutation relation, $\left[\hatX,\hatP\right] = i/2$, is used.
Using \eref{Eq_X_P}, we can write the Hamiltonian in \eref{HeisenH_ada}
in terms of the continuous variables $\hatX$ and $\hatP$, which we collectively denote
as $q$ for conciseness as follows:
\ba
\hat{H} =  \sum\limits_{k,l =1, k<l}^{N}
       \Bigg(&-&\frac{J_{kl}^x + J_{kl}^y}{4}~C_{k_1k_2l_1l_2}~(\hatq_{k_1}~\hatq_{k_2}~\hatq_{l_1}~\hatq_{l_2}
                                               ) \nonumber \\
             &-&\frac{J_{kl}^x - J_{kl}^y}{4}~F_{k_1k_2l_1l_2}~(\hatq_{k_1}~\hatq_{k_2}~\hatq_{l_1}~\hatq_{l_2}
                                               ) \nonumber \\
       &-& J^y_{kl} (q_{k_1}^2\hatq_{l_1}^2 + \hatq_{k_2}^2\hatq_{l_2}^2
                     -\hatq_{k_1}^2\hatq_{l_2}^2  - \hatq_{k_2}^2\hatq_{l_1}^2) \Bigg) \nonumber \\
       &-& \frac{B_0}{2} \sum\limits_{k=1}^{N} \hatq_k^2~.
\label{HeisenH_q}
\ea
In the equation above we have used the following convention to denote the
indices ($k, l$) and the spin orientation $(\uparrow, \downarrow)$: $k_1 :=
k\uparrow$, $k_2 := k\downarrow$, $l_1 := l\uparrow$, $l_2 := l\downarrow$. When
transforming from creation/annihilation operators to $\hatX$ and $\hatP$ variable, we
find that $\hatq_{k_1}~\hatq_{k_2}~\hatq_{l_1}~\hatq_{l_2}$ is non-zero only for 8 combinations
of $\hatX$ and $\hatP$ variable. These non-zero terms are shown in
Table-\ref{Table_terms}, for which $C=\pm1$ and $F=\pm1$, for various
combinations of the position and momentum variables $\hatX$ and $\hatP$ and the indices
$k_i$ and $l_i$.

\section{CV Gate decomposition}
\label{sec_3}
Typically the total Hamiltonian is expressed in
terms of summation of various terms. In our case, the Hamiltonian in
\eref{HeisenH_q} is composed of two quartic terms with coefficients $C$ and
$F$, product of quadratic terms and one quadratic term. To simulate the time
evolution operator, $e^{i\ \hatH\ t}$, where $\hatH= \sum_{j} \ \hatH_{j}$ is a
sum of operators with $\hatH_j$ denoting the quartic and quadratic terms, one can
use the Lie-Trotter product formula \cite{Suzuki:1976be,COHEN198255}. Trotterization
approximates the time
evolution operator as a series of $K$ time steps of size ${t}/{K}$,
during which, the non-commutativity of the Hamiltonian terms, $\hatH_{j}$, is
neglected. Precisely,
\begin{equation}
\label{Eq_Tott}
e^{i t\ \sum\limits_{j} ^{N}\hatH_{j}} \approx ( \prod\limits_{j}^{N} \ e^{i\frac{t}{K} \ \hatH_{j}})^{K} + O \left( \frac{N^{2} t^{2}\ \Gamma^{2}}{K}\right)
\end{equation}
where $\Gamma =  \max\limits_{j}\  \Vert \hatH_{j}\Vert$ is the largest Hamiltonian
norm. The last term accounts for the error in this approximation and it vanishes
as $K \rightarrow \infty$. However, infinite product terms would be equivalent
to an infinitely long circuit depth which is not feasible in practice.
Therefore, one needs to find a trade-off so that 
$K$ is large enough to achieve a reasonable accuracy, and at the same time, small 
enough to avoid exploiting the number of gates in the quantum circuit. 

Using \eref{Eq_Tott}, the time evolution operator of $\hatH$ in
\eref{HeisenH_q} is approximated as,

\begin{eqnarray}
e^{ i~t\ \hatH} \approx \Big[\prod_{k<l} \prod_{\hat{q}\in \hat{X}, \hat{P}\}} &&
\Big( e^{i \ \frac{t g_{kl}}{2 K} \ (\hatq_{k_1}~\hatq_{k_2}~\hatq_{l_1}~\hatq_{l_2})} \\ \nonumber
&\times &  \ e^{i \ \frac{tJ^{z}_{kl} }{K} \ (\hatq_{k_1}^2\hatq_{l_1}^2 + \hatq_{k_2}^2\hatq_{l_2}^2
                     -\hatq_{k_1}^2\hatq_{l_2}^2  - \hatq_{k_2}^2\hatq_{l_1}^2)}\\ \nonumber
&\times & e^{i \frac{B_{0} t}{K} \ \hat{q}_{k}^2} \Big) \Big] ^{K},
\end{eqnarray}

where the coefficients $g_{kl} := \frac{1}{2} ( C_{kl} \ (J_{kl}^{x} +
J^{y}_{kl}) + F_{kl}\  (J_{kl}^{x} - J^{y}_{kl}))$ and $C$ and $F$ are given explicitly in 
Table-\ref{Table_terms}. Similarly to \eref{HeisenH_q} we have used the
notation: $k_1 := k\uparrow$, $k_2 := k\downarrow$, $l_1 := l\uparrow$, $l_2 := l\downarrow$ \\

\subsection{Universal optical gates}

In order to optically implement the above time evolution operator, one needs to
decompose each term as a sequence of universal optical gates, defined as
follows \cite{RevModPhys.84.621,Kalajdzievski2019ExactGD}:
\ba
  R &:=& e^{i \alpha\  \hatX}, \nonumber \\  
  G &:=& e^{i \alpha {\hatX}^2}, \nonumber \\ 
  V &:=& e^{i \alpha {\hatX}^3}, \nonumber \\  
  \mathcal{F} &:=& e^{i \frac{\pi}{2} ( {\hatX}^2 + {\hatP}^2)},  \nonumber \\
  C_{z} &:=& e^{i \alpha \hatX  {\hatX^\prime}}. 
\ea

The above set consists of single mode gates namely rotation gate ($R$), Gaussian gate ($G$), 
cubic phase gate ($V$) and Fourier transform gate ($\mathcal{F}$) and two modes gate
 control phase gate ($C_{z}$). As discussed later in this section,
we need two additional gates for a complete decomposition of the time evolution operator: 
\begin{enumerate}
\item quartic gate, $Q_k(\alpha) =e^ {i \alpha\ \hatX_k^4}$
\item the translating or the shift gate, $T_{k} ({A})= e^{ i  \alpha\ \hatP_{k}~ {A}}$, where $A$ is the shift parameter.
\end{enumerate}
Let us show how to implement $Q$ and $T$ using the universal gate set.  
For an arbitrary operator function $f(\hatX)$, it is straightforward to show that
$T_k({A})\ f({\hatX}_k)\ T_{k}(A)^\dagger = f({\hatX}_k+{A})$ and $\mathcal{F} \
f({\hatX})\ \mathcal{F}^{\dagger} =f({\hatP})$. Furthermore, using the identity relation
\be
x^{4}_{2} = (x_{1} +x_{2}^{2})^2 - x_{1}^2 - 2\ x_{1} \ x_{2}^2,
\ee
the quartic gate $Q(\alpha)$ can be decomposed to: 
\ba
Q_{2}(\alpha)&=& e^{i\ \alpha \ {\hatX}^{4}_{2}}\\ \nonumber
&=& e^{ i  \ \hatP_{1} ~ {\hatX}^{2}_{2}} \ e^{i  \alpha\ \hatX_{1}^{2}}\ e^{- i  \ \hatP_{1}
~{\hatX}^{2}_{2}}\ \ e^{- i \alpha \ {\hatX}_{1}^2} \ e^{-2 i \alpha \ {\hatX}_{1} ~{\hatX}_{2}^{2}} \\ \nonumber
&=& T_{1}({\hatX}^{2}_{2}) \ G_{1}(\alpha) \ T^{\dagger}_{1}({\hatX}^{2}_{2}) \
G_{1}(-\alpha) \ \mathcal{F}_{1} \ T_{1}(-\alpha\  {\hatX}^{2}_{2}) \
\mathcal{F}_{1}^{\dagger}.
\ea
Similarly to the quartic gate $Q_k$, the shift gate $T_k$ can also be decomposed in
terms for the universal gate set. As discussed in Ref. \cite{Kalajdzievski2018ContinuousvariableGD}:
\begin{eqnarray}
e^{i\ P_{1} ~ \hatX_{2}} &=& \mathcal{F}_{1} \ C_{z} \ \mathcal{F}_{1}^{\dagger}\\
e^{i3\alpha^{2}\ \ \hatP_{1}  \hatX_{2}^{2}} &=& e^{i \hatP_{1}^3} \ e^{ -i \alpha \ {\hatX}_{1} {\hatX}_{2}}\ e^{- i \hatP_{1}^3}  \ e^{ -2i \alpha \ {\hatX}_{1} {\hatX}_{2}} \ e^{i \hatP_{1}^3} \nonumber  \\
&& \times \ e^{ i \alpha \ {\hatX}_{1} {\hatX}_{2}}\ e^{- i \hatP_{1}^3}\ e^{  2i \alpha \ {\hatX}_{1} {\hatX}_{2}} \ e^{i \frac{3}{4} \alpha^{3} \hat{X}_{2}^{3}}\\ 
 &=& V'_{1}\ C_{z}(-\alpha) \ V'^{\dagger}_{1}\ C_{z}(-2 \alpha) \ V'_{1} \nonumber \\
&& \times \ C_{z}(\alpha) \  V'^{\dagger}_{1} \ C_{z}(2\alpha) \ V_{2}(\frac{3}{4}
\alpha^{3}),
\label{quartic_decomp}
\end{eqnarray}
\noindent where $V'= \mathcal{F} \ V \mathcal{F}^{\dagger}$. In principle, one can find a
general decomposition relation for $e^{i3\alpha^{2} \hatP_{1}  \hatX_{2}^{n}}$
\cite{Kalajdzievski2019ExactGD}. However, up to the second
order is enough for our purposes.

Note that for implementing every quartic
operator, $Q$, three second-order shift gate are required, and, for each
of them, five cubic phase gates ($V$) and three phase control gates ($C_{z}$)
are required. Thus, in total, each $Q$ costs us 15 cubic phase gates ($V$) and 9
control phase gates ($C_{z}$). Now that we have understanding of the
decomposition of quartic gate and shift gate, we elaborate on the
decomposition of time evolution operator for the Hamiltonian in
\eref{HeisenH_q}.

\subsection{Decomposing the spin Hamiltonian}
In the following, we consider each individual term of the Hamiltonian in
\eref{HeisenH_q} and discuss their decomposition in terms of continuous
variable gates.
\begin{itemize}

\item {\bf Four mode quartic terms}: Let consider the first and second term in the
Hamiltonian which are of the form $\hatq_{k_1}~\hatq_{k_2}~\hatq_{l_1}~\hatq_{l_2}$. These terms
are quartic in order and contain products of four modes. In order to implement 
this in terms of the gates of the form $X^4$ we can use the following identity:
\begin{eqnarray}
\label{Eq_ID_1}
\hatq_{k_1}~\hatq_{k_2}~\hatq_{l_1}~\hatq_{l_2} = 
    \frac{1}{192} &\Big[& (\hatq_{k_1}+ \hatq_{k_2} + \hatq_{l_1} + \hatq_{l_2})^4  \nonumber \\ 
                 &+& (\hatq_{k_1} - \hatq_{k_2} - \hatq_{l_1} - \hatq_{l_2})^4 \nonumber \\
                 &-& (\hatq_{k_1} - \hatq_{k_2} + \hatq_{l_1} + \hatq_{l_2})^4 \nonumber \\
                 &-& (\hatq_{k_1} + \hatq_{k_2} - \hatq_{l_1} + \hatq_{l_2})^4 \nonumber \\
                 &-& (\hatq_{k_1} + \hatq_{k_2} + \hatq_{l_1} - \hatq_{l_2})^4 \nonumber \\
                 &+& (\hatq_{k_1} - \hatq_{k_2} - \hatq_{l_1} + \hatq_{l_2})^4 \nonumber \\
                 &+& (\hatq_{k_1} - \hatq_{k_2} + \hatq_{l_1} - \hatq_{l_2})^4 \nonumber \\
                 &+& (\hatq_{k_1} + \hatq_{k_2} - \hatq_{l_1} - \hatq_{l_2})^4 \Big].
\label{quartic}
\end{eqnarray}
Note that each of these quartic gates is in the form of
summation (with + or - signs) of $\hatq_{k_1},~\hatq_{k_2},~\hatq_{l_1}$ and $\hatq_{l_2}$, e.g.
the first term is $(\hatq_{k_1}+ \hatq_{k_2} + \hatq_{l_1} + \hatq_{l_2})^4$. These terms can be
obtained by starting with a quartic gate corresponding to $\hatq_{k_1}^4$ followed
by shifting it three times as follows:
\ba
 e^{i\alpha (\hatq_{k_1}+ \hatq_{k_2} + \hatq_{l_1} + \hatq_{l_2})^4} &=&  
       T_1(\hatq_{k_2}) T_1(\hatq_{k_3})  T_1(\hatq_{k_4}) ~ e^{i\alpha \hatq_{k_1}^4} \nonumber \\ 
       && T_1(\hatq_{k_2})  T_1(\hatq_{k_3})  T_1(\hatq_{k_4}).
\ea
Using similar expressions we can obtain the quartic gate implementation of the
each term on the right hand side of \eref{quartic}. Hence, in order to implement 
each of the quartic terms on the right hand
side of \eref{quartic}, we need 6 first order shift gate, in
addition to the three second order shift gates. This amounts to a total of
$8\times6=48$ first order and $8\times3=24$ second order shift gate. 

As an alternative to \eref{quartic}, we can expand the product of four modes
using the following identity:
\begin{eqnarray}\nonumber
\hatq_{k_1}~\hatq_{k_2}~\hatq_{l_1}~\hatq_{l_2} = \frac{1}{24} &\Big[& (\hatq_{k_1}+ \hatq_{k_2} + \hatq_{l_1} + \hatq_{l_2})^{4} \nonumber \\
&-& \sum\limits_{l\neq m\neq n} \ (\hatq_l+ \hatq_{m} + \hatq_{n})^{4} \nonumber \\
&+& \sum\limits_{l\neq m } \ (\hatq_l+ \hatq_{m})^{4} - \sum\limits_{l} \ \hatq_l^{4}\
\Big].
\end{eqnarray}
Here, $l,m,n \in \{{k_1},~{k_2},~{l_1},~{l_2}\}$.
For this implementation, $(1 \times 6 + 4 \times 4 + 6 \times 2 + 4) = 38$
first-order shift gates and $ 3 \times (1+4+6+4) = 45$ second-order
shift gates. Therefore, by choosing the first identity, we need much
fewer (approximately half as many) second order shift gates while the
number of first order shift gates are comparable for the two identities.
Therefore, the first identity in \eref{quartic} requires fewer cubic phase gates, 
$V$ and control phase gates, $C_{z}$.  

Note that $\hat{q}$ in \eref{HeisenH_q} can be either
position or momentum operator. We assume that all momentum operators can be
converted to the position operator using a Fourier transform gate, i.e.
$f(\hat{P})= \mathcal{F} \ f(\hat{X}) \ \mathcal{F}^{\dagger}$. Therefore, by
considering the first identity relation in \eref{Eq_ID_1}, the decomposition
of four-mode position operator is
\begin{widetext}
\begin{eqnarray}\nonumber
e^{i \alpha \ (\hat{X}_{i} \hat{X}_{j} \hat{X}_{\mu}  \hat{X}_{\nu})} &=&
\prod\limits_{\vec{s}} \  S_{\vec{s}}\big[\ T_{i}(\hat{X}_{\nu}) \
T_{i}(\hat{X}_{\mu}) \ T_{i}(\hat{X}_{j}) \ {e^{- i \frac{\alpha}{192}\hat{X}_{i}^4}}\ T_{i}^{\dagger}(\hat{X}_{j}) \  T_{i}^{\dagger} (\hat{X}_{\mu})\ T_{i}^{\dagger} (\hat{X}_{\nu})\big]\\
\end{eqnarray}
\end{widetext}
where the $S$ operators determines the sign of the elements in $(X_{j}, X_{\mu},
X_{\nu})$ by spanning $\vec{s} \in\{ (+, +, +), (-,-,-), (+,-,+), (+, +, -), (-,
+, +)\}$. Hence, four modes terms in the Hamiltonian can be written in terms of
quartic operators. The quartic operators can further be implemented
in terms of cubic phase gates and control phase gates as shown in 
\eref{quartic_decomp}. 

\item {\bf Two mode quartic terms}: The third term in \eref{HeisenH_q}
is also quartic, however, it appears as a product of two quadratic terms
corresponding to two different modes, i.e., ${q}_{l}^{i}  {q}_{j}^{2}$. Thus, we use the identity relation
\begin{equation}
\hatq_{i}^2\ \hatq_{j}^2 = \frac{1}{12}\ \{ (\hatq_{i} +\hatq_{j})^4 + (\hatq_{i} -\hatq_{j})^4 -2\ \hatq_{i}^2 - 2\ \hatq_{j}^2\},
\end{equation}
which leads to the following decomposition in terms of quartic gates:
\begin{eqnarray}
e^{i \alpha\ (\hat{X}_{l}^2 \hat{X}_{m}^2)} &=&  T_{l}(\hat{X}_{m}) \ e^{i \frac{\alpha}{12} \hat{X}_{l}^4}\ T_{l}^{\dagger}(\hat{X}_{m})  \\ \nonumber
 &\times &  T_{l}(-\hat{X}_{m}) \ e^{i \frac{\alpha}{12} \hat{X}_{l}^4}\ T_{l}^{\dagger}(-\hat{X}_{m}) \\ \nonumber
 &\times & e^{-2i \alpha \ \hat{X}_{l}^2}  \times e^{-2i \alpha \ \hat{X}_{m}^2}. 
\end{eqnarray}
Therefore, for each of the terms we need two quartic operators, four shift
gates and two quadratic operators. The quartic operators can be implemented
in terms of cubic phase gates and control phase gates as shown in
\eref{quartic_decomp}. The quadratic operators can be easily implement using
Gaussian gates.
Like before, for those terms that $\hat{q}=\hat{P}$ operator, we need to first apply a Fourier transform operator and turn it to an $\hat{X}$ operator and and then use the above decomposition.\\

\item {\bf Single mode quadratic terms}: The last term of the Hamiltonian, i.e.,
$\hat{q}^{2}$, is due to interaction with external magnetic field and it can be simply 
implemented by using a Gaussian gate, i.e., $G(\alpha) = e^{i t \alpha \hat{X}^2}$.
\end{itemize}

Using the decomposition discussed above, all the terms of the Hamiltonian in
\eref{HeisenH_q} can be implemented using universal CV gates. 



\section{Hybrid Quantum Simulation}
\label{sec_4}
Computation of the expectation value of the Hamiltonian using the CV gate
decomposition discussed in the previous section requires non-linear gates and
a fault tolerant CV quantum computer which is not expected to be available in
the near future. However, it has been shown that the noisy intermediate scale
quantum devices (NISQ) can potentially be used to implement hybrid quantum
algorithm while the full fledged quantum computer is under development
\cite{Peruzzo_2014,McClean_2016,PhysRevX.6.031007,Linke_2017,Kandala_2017,PhysRevLett.120.210501,McCaskey2019QuantumCA}.
In this section, we discuss a hybrid method to estimate the expectation value of
the spin Hamiltonian discussed here using an optical continuous variable NISQ
device.

Recall that, in order to implement a CV version of variational quantum
eigensolver algorithm, we need to estimate the expectation value of each terms
of the Hamiltonian, i.e. $\langle H \rangle = \sum\limits_{j=1} \ \ \langle
H_{j} \rangle$. The quadratic terms can be computed using photon counting
measurement of each mode: terms with $\langle q_i^{2} \rangle$ and 
$\langle q^{2}_{i} q^{2}_{j} \rangle$ are given by
photon number measurement $\langle\hat{\mathbf{n}}_{i}\rangle$ and two detector
correlation measurement $\langle\hat{\mathbf{n}}_{i}
\hat{\mathbf{n}}_{j}\rangle$. For the quartic terms containing four modes,
$\hatq_{k_1}~\hatq_{k_2}~\hatq_{l_1}~\hatq_{l_2}$, which would require non-linear gates and a
fault tolerant quantum computer, we present the following proposal where we
use the so-called Gaussian Boson sampling, a continuous variable NISQ device.

\subsection{Using Gaussian Boson sampling (GBS)}
The Gaussian Boson sampling device measures the system in the photon number basis (e.g. Bosonic Fock states of four modes are, $|n_{k_1} n_{k_2} n_{l_1} n_{l_2}\rangle$) and samples from the following distribution 
\begin{equation} \text{Pr}(\bar{m}) = \frac{1}{\sqrt{\bar{m}! \ \text{det}\tilde{\sigma}_{Q} }} \ \text{Haf}(A_{S})
\end{equation}
where $\bar{m} =  m_{1} m_{2}\dots m_{M}$ is a string of photon numbers detected
in $M$ modes and $\tilde{\sigma}_{Q} = \bm{\sigma}_{A} + \mathbb{1}_{2M}/2$ with
$\bm{\sigma}_{A}$ being the covariance matrix of the physical modes of the GBS
device \cite{gbs1,gbs2}. The $A_{S}$ is a submatrix of a matrix $A$
where $S$ is determined by the detected photon string, $\bar{m}$.
The matrix $A$ is related to the covariance matrix of the physical modes via 
\begin{equation}
\label{Eq_A_matrix}
A = X_{2M} \ [ \mathbb{I}_{2M}- (\bm{\sigma}_{A} + \mathbb{I}_{2M}/2)^{-1}]
\end{equation}

Let $\vec{\textbf{q}} =(\hatq_{1}, \hatq_{2}, \dots, \hatq_{m})$ be a random vector that
follows a normal distribution $ \vec{\textbf{q}} \sim N(\mu, \bm{\Sigma})$,
where $\mu$ is the mean and $\bm{\Sigma}= \{ \langle \hatq_{i}, \hatq_{j} \rangle \}$ is
the covariance matrix. For the case of zero mean, $\mu=0$, with even number of
modes, $m$, the mean value of the products is \cite{wick},

\begin{eqnarray}
\langle \hatq_{1}, \hatq_{2}, \dots, \hatq_{m} \rangle &=& \text{Haf}(\bm{\Sigma}).\\ \nonumber
\end{eqnarray}
The above theorem relates the expectation value of each individual term of the transverse Hamiltonian to the Hafnian of the covariance matrix.

If we use a direct mapping from the modes of interests, $\hatq_{i}$ to the physical
modes of the GBS device (i.e., if we choose $ \bm{\Sigma}= \bm{\sigma}_{A}$),
the GBS device can obtain an estimation of $\text{Haf}(A)$ after post selection
on $P(1,1, \dots, 1)$ where one photon is detected in all modes. However, the
quantity of our interest is $\text{Haf}(\bm{\Sigma})$ not $\text{Haf}(A)$. Since
the relation between $A$ and the covariance of the physical system in
\eref{Eq_A_matrix} is not linear, we expect that extracting
$\text{Haf}(\bm{\Sigma})$ from   $\text{Haf}(A$) is very non-trivial. Therefore,
the direct mapping where $\bm{\sigma}_{A} = \bm{\Sigma}$ is not suitable. \\

 One can consider engineering the GBS circuit such that $A=\bm{\Sigma}$. However, in general, such a matrix $A$ does not relate to valid covariance matrix to represent a physical system. Let $A=\begin{pmatrix}
A_{11} & A_{12} \\
A_{21} & A_{22}
\end{pmatrix}$ be a $2M \times 2M$ block diagonal symmetric matrix. According to
\cite{kamil1}, one of the requirements that $A$ must satisfy in order to map to a valid covariance matrix of a Gaussian state is $[A_{11}, \ A_{12} ] =0$ and $A_{12} \geq 0$. On the other hand, a covariance matrix $\bm{\Sigma}$ with $2M \times 2M$ dimension and is given by
\begin{eqnarray}
\bm{\Sigma} &=& \begin{pmatrix}
C & D \\
D^* & C
\end{pmatrix}, \\ \nonumber
C_{ij} &=& \frac{1}{2} \langle \{\hat{b}_{i}, \hat{b}_{j}^{\dagger}\} \rangle , \\ \nonumber
D_{ij} &=& \langle \hat{b}_{i} \ \hat{b}_{j} \rangle
\end{eqnarray}

where $\hat{b}_{i}^{\dagger}, \hat{b}_{i}$ are Bosonic creation-annihiliation
operators. In general, $[C, D] \neq 0$, and therefore, if we consider $A=
\bm{\Sigma}$, a mapping between the modes of interest and the physical modes of
the GBS device may not exist for an arbitrary $\bm{\Sigma}$. However, it was
shown in Ref. \cite{kamil1} that by doubling the number of optical modes and considering $A=\begin{pmatrix}
\bm{\Sigma} & 0 \\
0 & \bm{\Sigma}
\end{pmatrix}$
 a valid covariance matrix $\bm{\sigma}_{A}$ exists. Thus, the quantity of our interest can be measured by

\begin{equation}
\langle {q}_{1} {q}_{2} {q}_{3} {q}_{4} \rangle= \text{Haf}(\bm{\Sigma}) = \sqrt{\text{Haf}(A)}
\end{equation}
We highlight that obtaining the above results has a few requirements which must
be considered for experimental implementation. First, the initial state needs to
be Gaussian with zero mean which can be easily implemented using a displacement
operator. Second, it is required to do a post-selection on the measurement
outcome with specific photon pattern $\bar{m} = (1,1 \dots, 1)$.  Therefore, a
single photon detector is required which is not easy to implement and currently
an active field of research. Moreover, in order to implement a spin lattice with
$N$ sites, we require $4 N$ optical modes, a factor of two due to Bosonization and
another factor of two for estimating the Hafnian. 

\section{Conclusion}
\label{sec_5}
In this paper, we studied gate decomposition of  Ising
spin system in the continuous variable (CV) paradigm of quantum computing. 
We used the Jordan-Wigner transformation to transform the Hamiltonian of
transverse Ising model into a Bosonic Hamiltonian. The result was then used to
decompose the time evolution operator of Ising model in terms of universal gate
sets in continuous variable quantum computing. Multi-mode terms were analyzed in
details to obtain a compact expression for gate components. In addition to
application in quantum circuit model, we discussed how the current NISQ
devices such as Gaussian Boson sampling can be used to estimate ground states of Ising-type
Hamiltonian. The result can potentially be used for hybrid classical-quantum
algorithms such as variational quantum eigensolver.

Ising spin systems are  among the most studied problems in quantum
simulation. In addition to their wide use in understanding phase transition in
magnetic systems, it has been shown that many NP hard problems can be mapped to
an Ising model \cite{LucasIsingFormulation}. Therefore, solving an Ising model
on a quantum computer opens up new avenues to simulating NP-hard problems on
quantum computer.

\section*{Acknowledgements}
The authors would like to thank Leonardo Banchi for discussions. RA would like to thank support from Transformative Quantum Technologies, Waterloo and Xanadu Quantum Technologies Inc.


\begin{thebibliography}{49}

\bibitem{Lloyd96}
S.~{Lloyd}, ``{Universal Quantum Simulators}'', {\em Science} {\bfseries 273}
  (1996), no.~5278, 1073--1078.

\bibitem{AspuruGuzik2005SimulatedQC}
A.~Aspuru-Guzik, A.~D. Dutoi, P.~J. Love, and M.~Head-Gordon, ``Simulated
  quantum computation of molecular energies.'', {\em Science} {\bfseries 309
  5741} (2005) 1704--7.

\bibitem{WhitfieldetalQuantumSimulation}
J.~D. Whitfield, J.~Biamonte, and A.~Aspuru-Guzik, ``Simulation of electronic
  structure hamiltonians using quantum computers'', {\em Molecular Physics}
  {\bfseries 109} (2011), no.~5, 735--750,
  \href{http://xxx.lanl.gov/abs/https://doi.org/10.1080/00268976.2011.552441}{
  https://doi.org/10.1080/00268976.2011.552441}.

\bibitem{Babbush_2017}
R.~Babbush, D.~W. Berry, Y.~R. Sanders, I.~D. Kivlichan, A.~Scherer, A.~Y. Wei,
  P.~J. Love, and A.~Aspuru-Guzik, ``Exponentially more precise quantum
  simulation of fermions in the configuration interaction representation'',
  {\em Quantum Science and Technology} {\bfseries 3} Dec (2017) 015006.

\bibitem{Georgescu:2013oza}
I.~M. Georgescu, S.~Ashhab, and F.~Nori, ``{Quantum Simulation}'', {\em Rev.
  Mod. Phys.} {\bfseries 86} (2014) 153,
 \href{http://xxx.lanl.gov/abs/1308.6253}{ arXiv:1308.6253}.

\bibitem{feynman1982simulating}
R.~P. Feynman, ``Simulating physics with computers'', {\em International
  journal of theoretical physics} {\bfseries 21} (1982), no.~6/7, 467--488.

\bibitem{PhysRevLett.79.2586}
D.~S. Abrams and S.~Lloyd, ``Simulation of many-body fermi systems on a
  universal quantum computer'', {\em Phys. Rev. Lett.} {\bfseries 79} Sep
  (1997) 2586--2589.

\bibitem{PhysRevLett.83.5162}
D.~S. Abrams and S.~Lloyd, ``Quantum algorithm providing exponential speed
  increase for finding eigenvalues and eigenvectors'', {\em Phys. Rev. Lett.}
  {\bfseries 83} Dec (1999) 5162--5165.

\bibitem{Zalka_1998}
C.~Zalka, ``Simulating quantum systems on a quantum computer'', {\em
  Proceedings of the Royal Society of London. Series A: Mathematical, Physical
  and Engineering Sciences} {\bfseries 454} Jan (1998) 313–322.

\bibitem{Kassal_2008}
I.~Kassal, S.~P. Jordan, P.~J. Love, M.~Mohseni, and A.~Aspuru-Guzik,
  ``Polynomial-time quantum algorithm for the simulation of chemical
  dynamics'', {\em Proceedings of the National Academy of Sciences} {\bfseries
  105} Nov (2008) 18681–18686.

\bibitem{Brown_2010}
K.~L. Brown, W.~J. Munro, and V.~M. Kendon, ``Using quantum computers for
  quantum simulation'', {\em Entropy} {\bfseries 12} Nov (2010) 2268–2307.


\bibitem{Brush_67_IsingRMP}
S.~G. BRUSH, ``History of the lenz-ising model'', {\em Rev. Mod. Phys.}
  {\bfseries 39} Oct (1967) 883--893.

\bibitem{stanley1987introduction}
H.~Stanley, ``Introduction to phase transitions and critical phenomena'',
  Oxford University Press, 1987.

\bibitem{binney1992theory}
J.~Binney, J.~Binney, M.~Binney, N.~Dowrick, A.~Fisher, M.~Newman, B.~Fisher,
  and L.~Newman, ``The theory of critical phenomena: An introduction to the
  renormalization group'', Clarendon Press, 1992.

\bibitem{Barahona_1982}
F.~Barahona, ``On the computational complexity of ising spin glass models'',
  {\em Journal of Physics A: Mathematical and General} {\bfseries 15} oct
  (1982) 3241--3253.

\bibitem{Lucas_2014}
A.~Lucas, ``Ising formulations of many np problems'', {\em Frontiers in
  Physics} {\bfseries 2} (2014).

\bibitem{Xia2017ElectronicSC}
R.~Xia, T.~Bian, and S.~Kais, ``Electronic structure calculations and the ising
  hamiltonian.'', {\em The journal of physical chemistry. B} {\bfseries 122 13}
  (2017) 3384--3395.

\bibitem{RevModPhys.84.621}
C.~Weedbrook, S.~Pirandola, R.~Garc\'{\i}a-Patr\'on, N.~J. Cerf, T.~C. Ralph,
  J.~H. Shapiro, and S.~Lloyd, ``Gaussian quantum information'', {\em Rev. Mod.
  Phys.} {\bfseries 84} May (2012) 621--669.

\bibitem{RevModPhys.77.513}
S.~L. Braunstein and P.~van Loock, ``Quantum information with continuous
  variables'', {\em Rev. Mod. Phys.} {\bfseries 77} Jun (2005) 513--577.

\bibitem{PhysRevLett.82.1784}
S.~Lloyd and S.~L. Braunstein, ``Quantum computation over continuous
  variables'', {\em Phys. Rev. Lett.} {\bfseries 82} Feb (1999) 1784--1787.

\bibitem{PhysRevA.93.052304}
N.~Liu, J.~Thompson, C.~Weedbrook, S.~Lloyd, V.~Vedral, M.~Gu, and K.~Modi,
  ``Power of one qumode for quantum computation'', {\em Phys. Rev. A}
  {\bfseries 93} May (2016) 052304.

\bibitem{nielsen_chuang_2010}
M.~A. Nielsen and I.~L. Chuang, ``Quantum computation and quantum information:
  10th anniversary edition'', Cambridge University Press, 2010.

\bibitem{lidar_ising}
D.~A. Lidar and O.~Biham, ``Simulating ising spin glasses on a quantum
  computer'', {\em Phys. Rev. E} {\bfseries 56} Sep (1997) 3661--3681.

\bibitem{Santoro_2002}
G.~E. Santoro, ``Theory of quantum annealing of an ising spin glass'', {\em
  Science} {\bfseries 295} Mar (2002) 2427–2430.

\bibitem{Boixo_2013}
S.~Boixo, T.~Albash, F.~M. Spedalieri, N.~Chancellor, and D.~A. Lidar,
  ``Experimental signature of programmable quantum annealing'', {\em Nature
  Communications} {\bfseries 4} Jun (2013).

\bibitem{Boixo_2014}
S.~Boixo, T.~F. Rønnow, S.~V. Isakov, Z.~Wang, D.~Wecker, D.~A. Lidar, J.~M.
  Martinis, and M.~Troyer, ``Evidence for quantum annealing with more than one
  hundred qubits'', {\em Nature Physics} {\bfseries 10} Feb (2014) 218–224.

\bibitem{Johnson2011Quantum}
M.~W. Johnson, M.~H.~S. Amin, S.~Gildert, T.~Lanting, F.~Hamze, N.~Dickson,
  R.~Harris, A.~J. Berkley, J.~Johansson, P.~Bunyk, E.~M. Chapple, C.~Enderud,
  J.~P. Hilton, K.~Karimi, E.~Ladizinsky, N.~Ladizinsky, T.~Oh, I.~Perminov,
  C.~Rich, M.~C. Thom, E.~Tolkacheva, C.~J.~S. Truncik, S.~Uchaikin, J.~Wang,
  B.~Wilson, and G.~Rose, ``Quantum annealing with manufactured spins'', {\em
  Nature} {\bfseries 473} (2011), no.~7346, 194--198.

\bibitem{Whitfield_2012}
J.~D. Whitfield, M.~Faccin, and J.~D. Biamonte, ``Ground-state spin logic'',
  {\em EPL (Europhysics Letters)} {\bfseries 99} Sep (2012) 57004.

\bibitem{Biamonte2008RealizableHF}
J.~D. Biamonte and P.~J. Love, ``Realizable hamiltonians for universal
  adiabatic quantum computers'', {\em Physical Review A} {\bfseries 78} (2008)
  012352.

\bibitem{Cervera_Lierta_2018}
A.~Cervera-Lierta, ``Exact ising model simulation on a quantum computer'', {\em
  Quantum} {\bfseries 2} Dec (2018) 114.

\bibitem{Peruzzo_2014}
A.~Peruzzo, J.~McClean, P.~Shadbolt, M.-H. Yung, X.-Q. Zhou, P.~J. Love,
  A.~Aspuru-Guzik, and J.~L. O’Brien, ``A variational eigenvalue solver on a
  photonic quantum processor'', {\em Nature Communications} {\bfseries 5} Jul
  (2014).

\bibitem{McClean_2016}
J.~R. McClean, J.~Romero, R.~Babbush, and A.~Aspuru-Guzik, ``The theory of
  variational hybrid quantum-classical algorithms'', {\em New Journal of
  Physics} {\bfseries 18} Feb (2016) 023023.

\bibitem{PhysRevX.6.031007}
P.~J.~J. O'Malley, R.~Babbush, I.~D. Kivlichan, J.~Romero, J.~R. McClean,
  R.~Barends, J.~Kelly, P.~Roushan, A.~Tranter, N.~Ding, B.~Campbell, Y.~Chen,
  Z.~Chen, B.~Chiaro, A.~Dunsworth, A.~G. Fowler, E.~Jeffrey, E.~Lucero,
  A.~Megrant, J.~Y. Mutus, M.~Neeley, C.~Neill, C.~Quintana, D.~Sank,
  A.~Vainsencher, J.~Wenner, T.~C. White, P.~V. Coveney, P.~J. Love, H.~Neven,
  A.~Aspuru-Guzik, and J.~M. Martinis, ``Scalable quantum simulation of
  molecular energies'', {\em Phys. Rev. X} {\bfseries 6} Jul (2016) 031007.

\bibitem{Linke_2017}
N.~M. Linke, D.~Maslov, M.~Roetteler, S.~Debnath, C.~Figgatt, K.~A. Landsman,
  K.~Wright, and C.~Monroe, ``Experimental comparison of two quantum computing
  architectures'', {\em Proceedings of the National Academy of Sciences}
  {\bfseries 114} Mar (2017) 3305–3310.

\bibitem{Kandala_2017}
A.~Kandala, A.~Mezzacapo, K.~Temme, M.~Takita, M.~Brink, J.~M. Chow, and J.~M.
  Gambetta, ``Hardware-efficient variational quantum eigensolver for small
  molecules and quantum magnets'', {\em Nature} {\bfseries 549} Sep (2017)
  242–246.

\bibitem{PhysRevLett.120.210501}
E.~F. Dumitrescu, A.~J. McCaskey, G.~Hagen, G.~R. Jansen, T.~D. Morris,
  T.~Papenbrock, R.~C. Pooser, D.~J. Dean, and P.~Lougovski, ``Cloud quantum
  computing of an atomic nucleus'', {\em Phys. Rev. Lett.} {\bfseries 120} May
  (2018) 210501.

\bibitem{McCaskey2019QuantumCA}
A.~J. McCaskey, Z.~P. Parks, J.~Jakowski, S.~V. Moore, T.~Morris, T.~S. Humble,
  and R.~C. Pooser, ``Quantum chemistry as a benchmark for near-term quantum
  computers'', {\em npj Quantum Information} {\bfseries 5} (2019) 1--8.


\bibitem{Lenz_1920}
W.~Lenz, ``Beitr\"age zum verst\'andnis der magnetischen eigenschaften in
  festen k\"orpern'', {\em Physikalische Zeitschrift} {\bfseries 21} (1920)
  613--615.

\bibitem{Ising_1925}
E.~Ising, ``Beitrag zur theorie des ferromagnetismus'', {\em Z. Physik}
  {\bfseries 31} December (1925) 253--258.

\bibitem{jordan1}
P.~Jordan, ``Der zusammenhang der symmetrischen und linearen gruppen und das
  mehrk\"orperproblem'', {\em Zeitschrift f\"ur Physik} {\bfseries 7} no.~94,.

\bibitem{schwinger2}
J.~Schwinger, ``On angular momentum'', {\em Harvard Univ., Cambridge, MA
  (United States); Nuclear Development Associates, Inc. (US)}.

\bibitem{Suzuki:1976be}
M.~Suzuki, ``{Generalized Trotter's Formula and Systematic Approximants of
  Exponential Operators and Inner Derivations with Applications to Many Body
  Problems}'', {\em Commun. Math. Phys.} {\bfseries 51} (1976) 183--190.

\bibitem{COHEN198255}
J.~E. Cohen, S.~Friedland, T.~Kato, and F.~P. Kelly, ``Eigenvalue inequalities
  for products of matrix exponentials'', {\em Linear Algebra and its
  Applications} {\bfseries 45} (1982) 55 -- 95.

\bibitem{Kalajdzievski2019ExactGD}
T.~Kalajdzievski and J.~M. Arrazola, ``Exact gate decompositions for photonic
  quantum computing'', {\em Physical Review A} {\bfseries 99} (2019) 022341.

\bibitem{Kalajdzievski2018ContinuousvariableGD}
T.~Kalajdzievski, C.~Weedbrook, and P.~Rebentrost, ``Continuous-variable gate
  decomposition for the bose-hubbard model'', {\em Physical Review A}
  {\bfseries 97} (2018) 062311.

\bibitem{gbs1}
C.~S. Hamilton, R.~Kruse, L.~Sansoni, S.~Barkhofen, C.~Silberhorn, and I.~Jex,
  ``Gaussian Boson sampling'', {\em Physical Review Letters} {\bfseries 119}
  Oct (2017).

\bibitem{gbs2}
R.~Kruse, C.~S. Hamilton, L.~Sansoni, S.~Barkhofen, C.~Silberhorn, and I.~Jex,
  ``Detailed study of gaussian Boson sampling'', {\em Phys. Rev. A} {\bfseries
  100} Sep (2019) 032326.

\bibitem{wick}
G.~C.~Wick, {\em Physical Review} {\bf 80} 268 (1950)

\bibitem{kamil1}
K.~Brádler, P.-L. Dallaire-Demers, P.~Rebentrost, D.~Su, and C.~Weedbrook,
  ``Gaussian Boson sampling for perfect matchings of arbitrary graphs'', {\em
  Physical Review A} {\bfseries 98} Sep (2018).

\bibitem{LucasIsingFormulation}
A.~{Lucas}, ``Ising formulations of many np problems'', {\em Frontiers in
  Physics} {\bfseries 2} (2014) 5,  \href{http://xxx.lanl.gov/abs/1302.5843}{
  arXiv:1302.5843}.

\end{thebibliography}
\end{document}